%% file: ms_final.tex
\documentclass[12pt, preprint]{aastex}

\usepackage{natbib}
\usepackage{psfig}
\usepackage{amsfonts,amsmath}
\usepackage{lscape}

\newcommand{\ha}{H$\alpha$\ }
\newcommand{\ang}{$\AA$\ }
\newcommand{\etal}{{et~al.}\ }

\begin{document}

\title{Short-Term \ha Variability in M Dwarfs}

\author{
Khee-Gan~Lee \altaffilmark{1}, 
Edo~Berger \altaffilmark{2}, 
and Gillian~R.~Knapp \altaffilmark{1}
}

\altaffiltext{1}{Department of Astrophysical Sciences, Princeton
University, Peyton Hall, Ivy Lane, Princeton, NJ 08544 USA;
lee@astro.princeton.edu}

\altaffiltext{2}{Harvard-Smithsonian Center for Astrophysics, 60
Garden Street, Cambridge, MA 02138, USA; eberger@cfa.harvard.edu}

\begin{abstract} We spectroscopically study the variability of \ha 
emission in mid- to late-M dwarfs on timescales of $\sim 0.1-1$ hr as
a proxy for magnetic variability.  About 80\% of our sample exhibits
statistically significant variability on the full range of timescales
probed by the observations, and with amplitude ratios in the range of
$\sim 1.2-4$.  No events with an order of magnitude increase in
H$\alpha$ luminosity were detected, indicating that their rate is
$\lesssim 0.05$ hr$^{-1}$ (95\% confidence level).  We find a clear
increase in variability with later spectral type, despite an overall
decrease in H$\alpha$ ``activity'' (i.e., $L_{\rm H\alpha}/L_{\rm
bol}$).  For the ensemble of H$\alpha$ variability events, we find a
nearly order of magnitude increase in the number of events from
timescales of about 10 to 30 min, followed by a roughly uniform
distribution at longer durations.  The event amplitudes follow an
exponential distribution with a characteristic scale of ${\rm
Max(EW)/Min(EW)}-1\approx 0.7$.  This distribution predicts a low rate
of $\sim 10^{-6}$ hr$^{-1}$ for events with ${\rm Max(EW)/Min(EW)}
\gtrsim 10$, but serendipitous detections of such events in the past
suggests that they represent a different distribution.  Finally, we
find a possible decline in the amplitude of events with durations of
$\gtrsim 0.5$ hr, which may point to a typical energy release in
H$\alpha$ events for each spectral type ($E_{\rm H\alpha}\sim L_{\rm
H\alpha}\times t\sim {\rm const}$).  Longer observations of individual
active objects are required to further investigate this possibility.
Similarly, a larger sample may shed light on whether H$\alpha$
variability correlates with properties such as age or rotation
velocity.
\end{abstract}


\section{Introduction}

One of the primary indicators of magnetic heating and activity in low
mass stars is \ha chromospheric emission, which traces the presence of
gas at temperatures of $\sim 5000-10,000$ K.  In M dwarfs and later
spectral types \ha is a particularly prominent indicator since it is
more easily accessible than other chromospheric lines such as
\ion{Ca}{2} and \ion{Mg}{2}, which are located in the faint blue part
of the spectrum (e.g., \citealt{hawley96}).  Moreover, information on
the \ha line is readily available as a by-product of any spectroscopic
observations that are used to classify M dwarf properties, such as the
TiO band-head at 7050 \ang.  Thus, large samples of M dwarfs now exist
with measurements of \ha luminosity and its ratio relative to the
bolometric luminosity, $L_{\rm H\alpha}/L_{\rm bol}$ (commonly
referred to as the \ha ``activity'').

These samples have led to several important results concerning
chromospheric activity in low mass stars.  First, the fraction of
objects that exhibit \ha in emission increases rapidly from $\sim 5\%$
in the K5--M3 dwarfs to a peak of $\sim 70\%$ around spectral type M7,
followed by a subsequent decline in the L dwarfs
\citep{gizis00,west04,west08}.  Second, while the level of activity
increases with both rotation and youth in F--K stars, it reaches a
saturated value of $L_{\rm H\alpha}/L_{\rm bol}\approx 10^{-3.8}$ in
M0--M6 dwarfs, followed by a rapid decline to $L_{\rm H\alpha}/L_{\rm
bol} \approx 10^{-5}$ by spectral type L0
\citep{hawley96,gizis00,west04}.  Third, a rotation-activity relation
is observed in spectral types earlier than $\sim {\rm M7}$, such that
essentially all objects with $v\sin(i)\gtrsim 5$ km s$^{-1}$ exhibit
saturated H$\alpha$ activity.  However, late-M and L dwarfs exhibit
reduced activity even at high rotation rates, $v\sin (i)\gtrsim 10$ km
s$^{-1}$ \citep{mb03,rb08}.  Finally, a small fraction ($\lesssim
5\%$) of late-M and L dwarfs have been serendipitously observed to
exhibit H$\alpha$ flares that reach the saturated emission levels
found in the earlier-M dwarfs \citep{lieb03}.

In this paper we focus on the last point (H$\alpha$ variability) in
the spectral type range M3.5--M8.5, which encompasses the peak of the
H$\alpha$ active fraction, the regime of saturated emission, and the
breakdown of the rotation-activity relation.  In this spectral type
range stars are also fully convective, indicating that the solar-type
$\alpha\Omega$ dynamo (e.g., \citealt{par55}) no longer operates.
Thus, studies of H$\alpha$ temporal variability provide additional
constraints on the magnetic dynamo mechanism.  While a large number of
objects in the mid- and late-M spectral type range have been included
in the studies outlined above (for example, $\sim 10^4$ objects in
\citealt{west08}), \ha variability was not a key aspect of the
observations and thus most of the flare and variability detections
have been serendipitous in nature.  As a result, no systematic results
on H$\alpha$ variability timescales and amplitudes are available in
the literature from controlled and uniform cadence observations of a
large sample of mid- and late-M dwarfs.  Indeed, the few existing
studies have only targeted H$\alpha$ variability in small samples of
early M dwarfs.  For example, \citet{bs78} studied 15 objects with
spectral types earlier than M4.5 and with a cadence of $\gtrsim 1$
day, while \citet{pce84} studied H$\alpha$ variability in only three
M3-M3.5 dwarfs.  \citet{grh02} studied a larger sample of mid- and
late-M dwarfs, but did not use a uniform cadence (or discuss what
their cadence was).  Finally, observations of a small number of late-M
and early-L dwarfs with durations of $\sim 8-10$ hr have revealed
periodic H$\alpha$ emission in two objects, tied to their rotation
period \citep{bgg+08,brp+09}.

Here we present spectroscopic observations of over 40 M3.5--M8.5
dwarfs designed to probe chromospheric variability on timescales of
about 5 min to 1 hr.  Focusing on the objects that exhibit H$\alpha$
emission, we find that nearly $80\%$ are variable over the full range
of timescales probed by our observations.  The outline of the paper is
as follows.  In \S\ref{sec:obs} we describe the observations and
measurements of the \ha equivalent widths and fluxes.  We study the
H$\alpha$ emission and its variability in \S\ref{sec:basicres}, and
finally discuss the observed trends and their implications in
\S\ref{sec:var}.

\section{Observations and Data Reduction}
 \label{sec:obs}

We targeted 43 well-studied M dwarfs in the spectral range M3.5 to
M8.5, selected from the samples of \citet{d98}, \citet{gizis02},
\citet{mb03}, \citet{pb06}, \citet{mb03}, \citet{cruz03}, and
\citet{crifo05}.  Given the observational setup (see below), we
selected targets at a distance of $\lesssim 25$ pc and with an
observed magnitude of $V\lesssim 20$ mag.  The properties of our
selected targets, including luminosities and rotation velocities when
available, are summarized in Table~\ref{tab:obstable}.  Robust ages
are not available for our objects, but the small nearby volume of the
sample indicates that few objects are expected to be very young.  The
majority of the objects in our sample have been previously shown to
have \ha emission, although only VB\,10 was known to exhibit flaring
in the chromospheric emission lines.

The observations were carried out using the Boller \& Chivens
Spectrograph mounted on the du Pont 2.5-m telescope at Las Campanas
Observatory, Chile, on two separate occasions: 2007 March 14--17 and
2007 September 12--17.  In all observations we used a slit width of
$1.5''$, matched to the average seeing conditions, and a 600 lines
mm$^{-1}$ grating blazed at 5000 \ang.  The spectral coverage extended
from about 3680 to 6850 \ang, designed to cover a wide range of the
hydrogen Balmer lines, the \ion{Ca}{2} H\&K doublet, and the
\ion{He}{1} lines.  The spectral resolution was about 5 \ang.

The March 2007 observations were carried out in good weather
conditions, with typical seeing of about $1.2''$, while the conditions
during the September 2007 observations were poorer, with strong winds
and a typical seeing of about $1.5''$.  We observed a total of 20 and
23 objects in the two runs, respectively, with individual exposures of
300 or 600 s (depending on the brightness of the object) and a total
observing time of about 1 hr per source.  Six of our objects were
observed more than once.  In total, we obtained over 600 individual
spectra spanning $\sim 3000$ min of total exposure time.  A log of the
observations is provided in Table~\ref{tab:obstable}.  The data were
reduced using standard routines in {\tt IRAF}, and the wavelength
calibration was performed using He-Ar arc lamp exposures.

To measure the \ha equivalent width (EW) in a uniform manner we fit a
second-order polynomial to the pseudo-continuum from 6540 to 6620 \ang
(excluding $\pm 10$ \ang around the \ha emission line).  The
equivalent widths were then determined by summing the area under the
\ha line.  The error on each equivalent width measurement includes the
noise in the spectrum over the same spectral range, as well as the
uncertainty in the continuum fit (using a $\chi^2$ statistic).

\section{\ha Equivalent Width Light Curves and Overall Variability}
\label{sec:basicres}

\begin{figure*}
\plotone{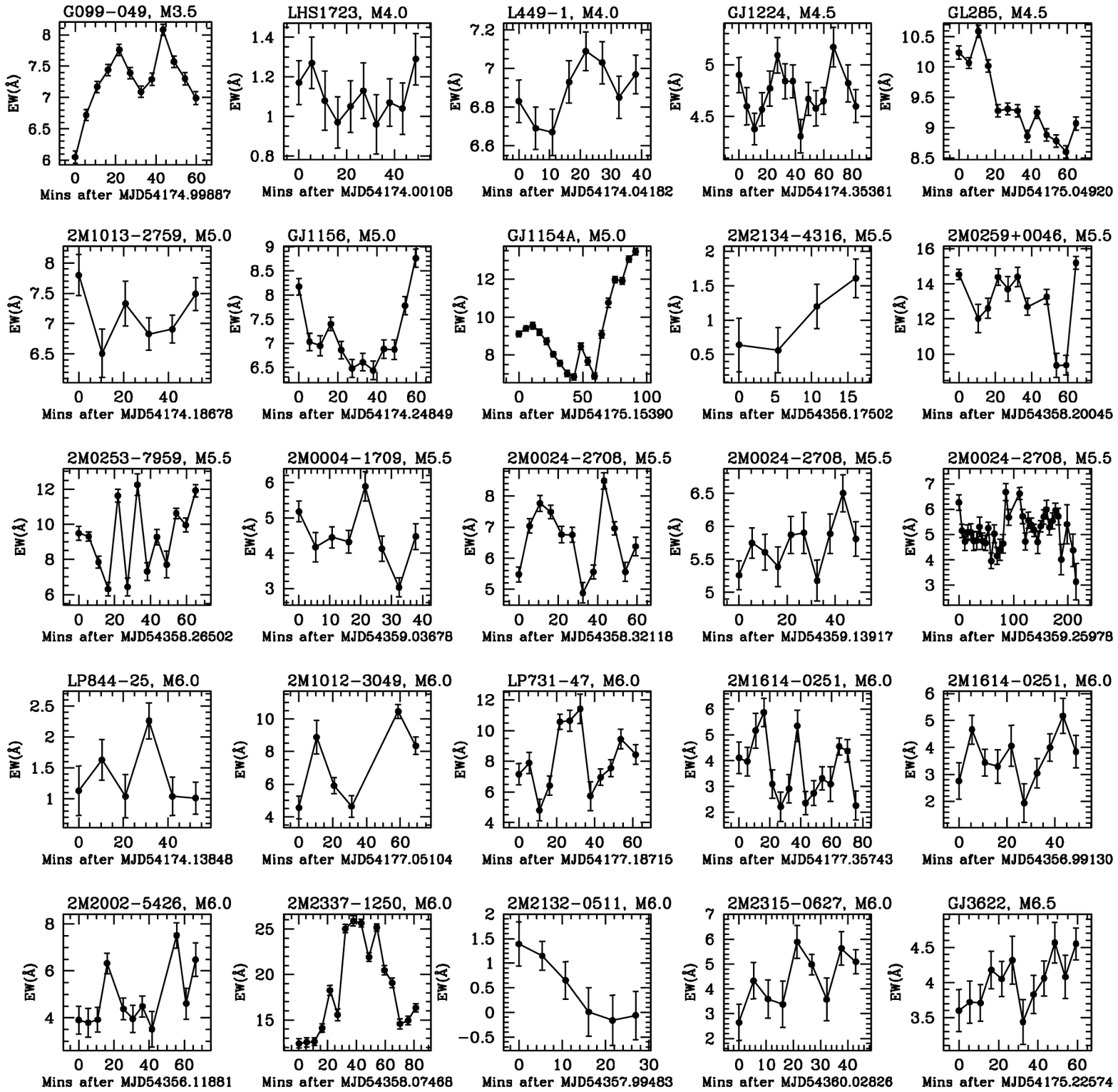}
\caption{\ha EWs plotted against time in minutes, for all observations
in our sample.  Note that some objects have several light curves from
repeated observations.
\label{fig:lcurve_all1}}
\end{figure*}

\begin{figure*}
\plotone{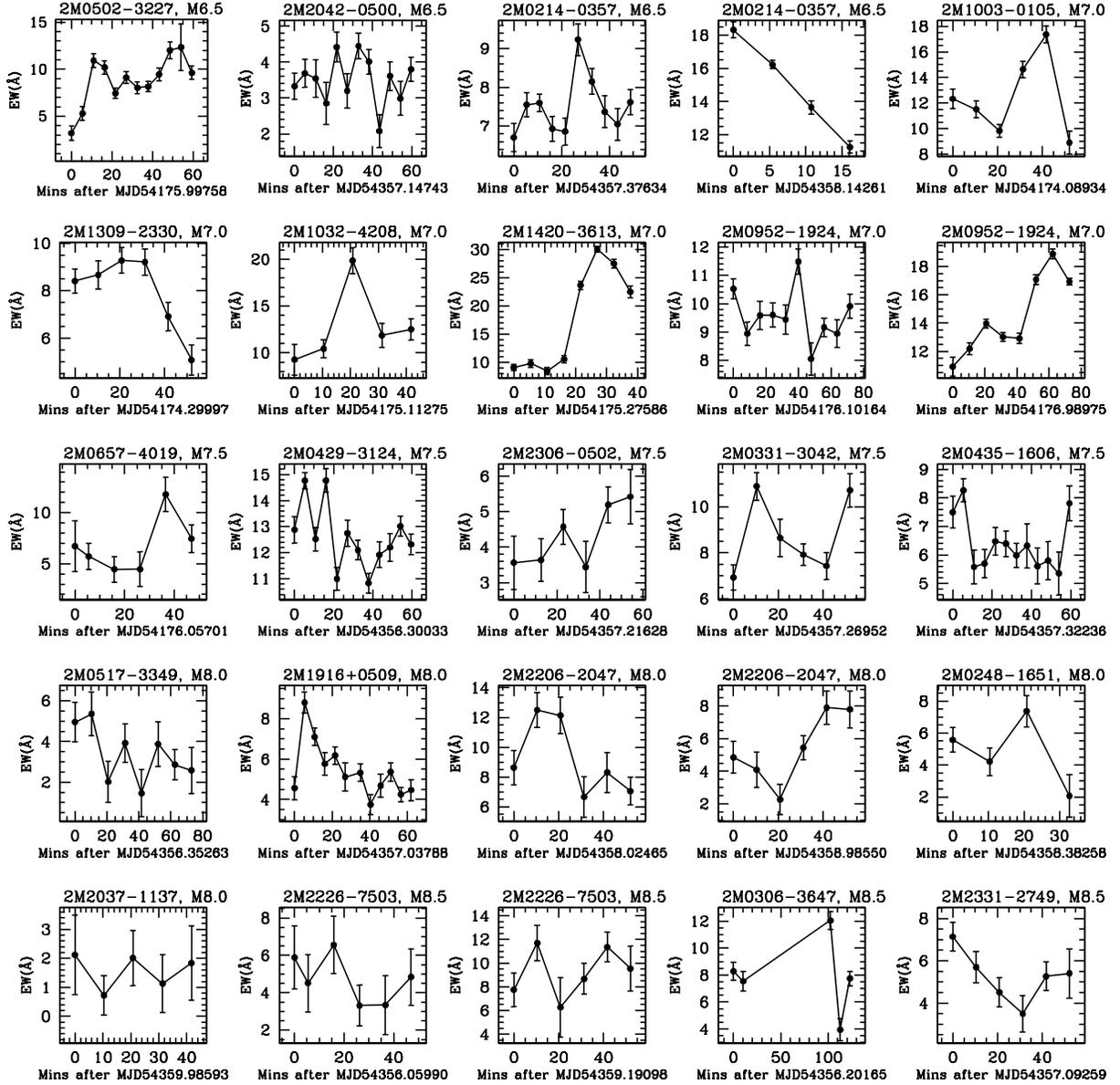}
\caption{Same as Figure~\ref{fig:lcurve_all1}. 
\label{fig:lcurve_all2}}
\end{figure*}

\begin{figure*}
\plotone{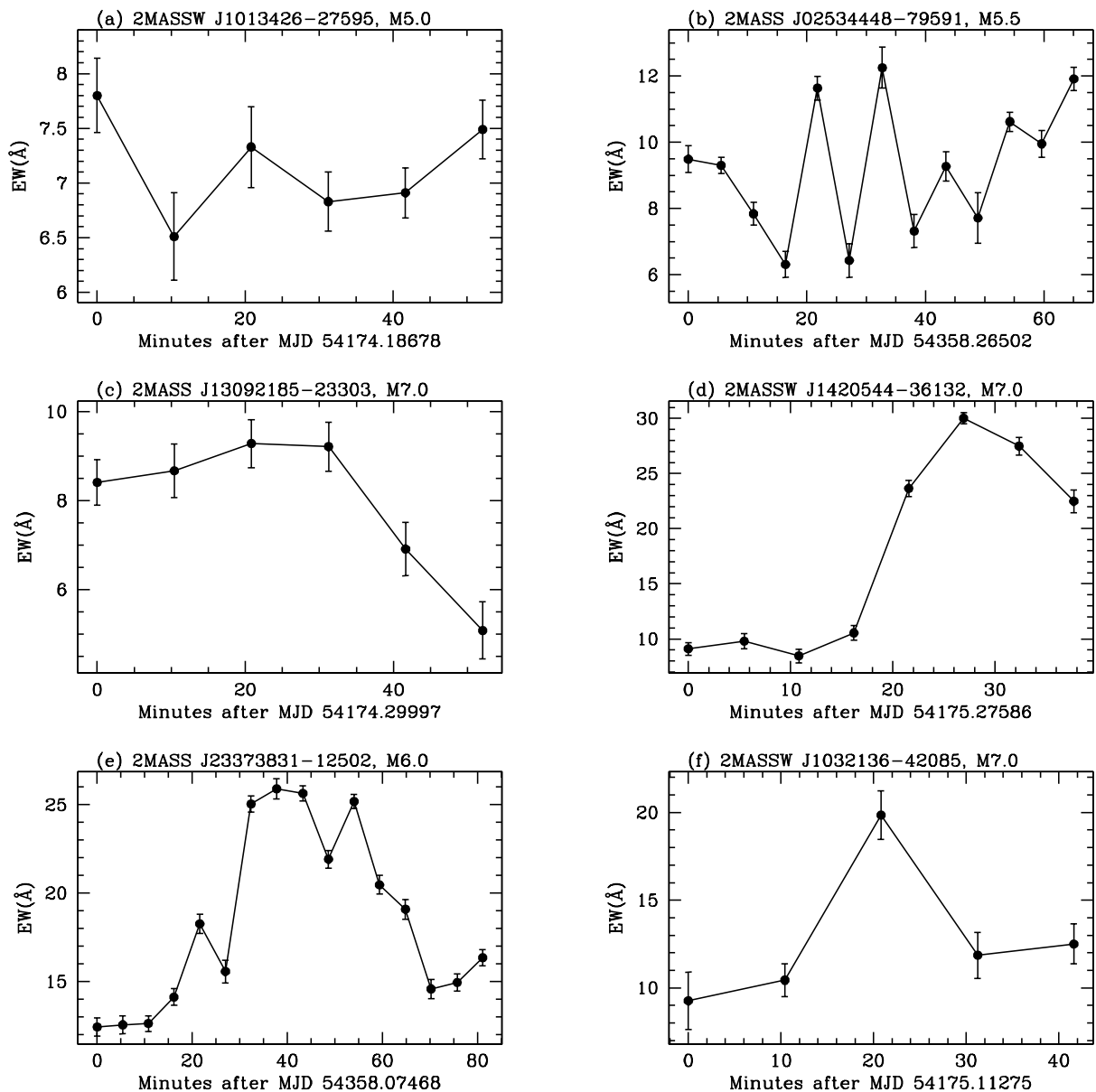}
\caption{Subset of Figures \ref{fig:lcurve_all1} and
\ref{fig:lcurve_all2} enlarged for discussion.
Figure~\ref{fig:lcurve}a is an example of a light curve consistent
with non-varying EW (see \S\ref{sec:basicres}).
Figures~\ref{fig:lcurve}b and \ref{fig:lcurve}c display examples of
objects with rapid and gradual variability, respectively.  The other
plots show 3 of the objects that displayed the greatest variability
during our observations.
\label{fig:lcurve}}
\end{figure*}

The \ha equivalent width light curves for all individual observations
are shown in Figures~\ref{fig:lcurve_all1} and \ref{fig:lcurve_all2}.
We highlight several individual sources in Figure~\ref{fig:lcurve}.
In Table~\ref{tab:vartable} we provide for each source a summary of
the median, minimum, maximum, and root-mean-square (RMS) scatter of
the \ha equivalent width (EW).

We determine whether a source is variable using a $\chi^2$ test.
Namely, we fit a straight line at constant EW through the light curve
and calculate the $\chi^2$ value for the best fit.  Nine of the
objects have light curves that are consistent with non-varying \ha
emission on timescales of $\sim 5-60$ min at a confidence level
greater than $95\%$ (one of them, 2M2226$-$7503, has two separate
observations consistent with non-varying emission).  These objects are
denoted with a `(C)' in the column of RMS values in
Table~\ref{tab:vartable}; Figure \ref{fig:lcurve}a shows the light
curve of one such object.  We thus conclude that only $\sim 20\%$ of
mid- to late-M dwarfs with \ha emission are non-variable on a
timescale of $\lesssim 1$ hr.

The rest of the sources ($\sim 80\%$) exhibit a wide range of
variability timescales and amplitudes, from rapid variations at the
minimum time resolution of our observations (e.g., the M5.5 object
2M0253$-$7959, Figure \ref{fig:lcurve}b) to slower variations that
span the entire observation (e.g., the M7 object 2M1309$-$2330, Figure
\ref{fig:lcurve}c).  Similarly, the variability amplitudes range from
about 1 \ang to over 20 \ang.  Naturally, the sensitivity to small
amplitude variations is a function of the signal-to-noise ratio, which
in turn depends on the brightness and hence spectral type.  The
typical variability amplitudes corresponds to fluctuations of about a
factor of two in the \ha luminosity.

\begin{figure*}
\begin{minipage}{2.00in}
\centerline{\psfig{file=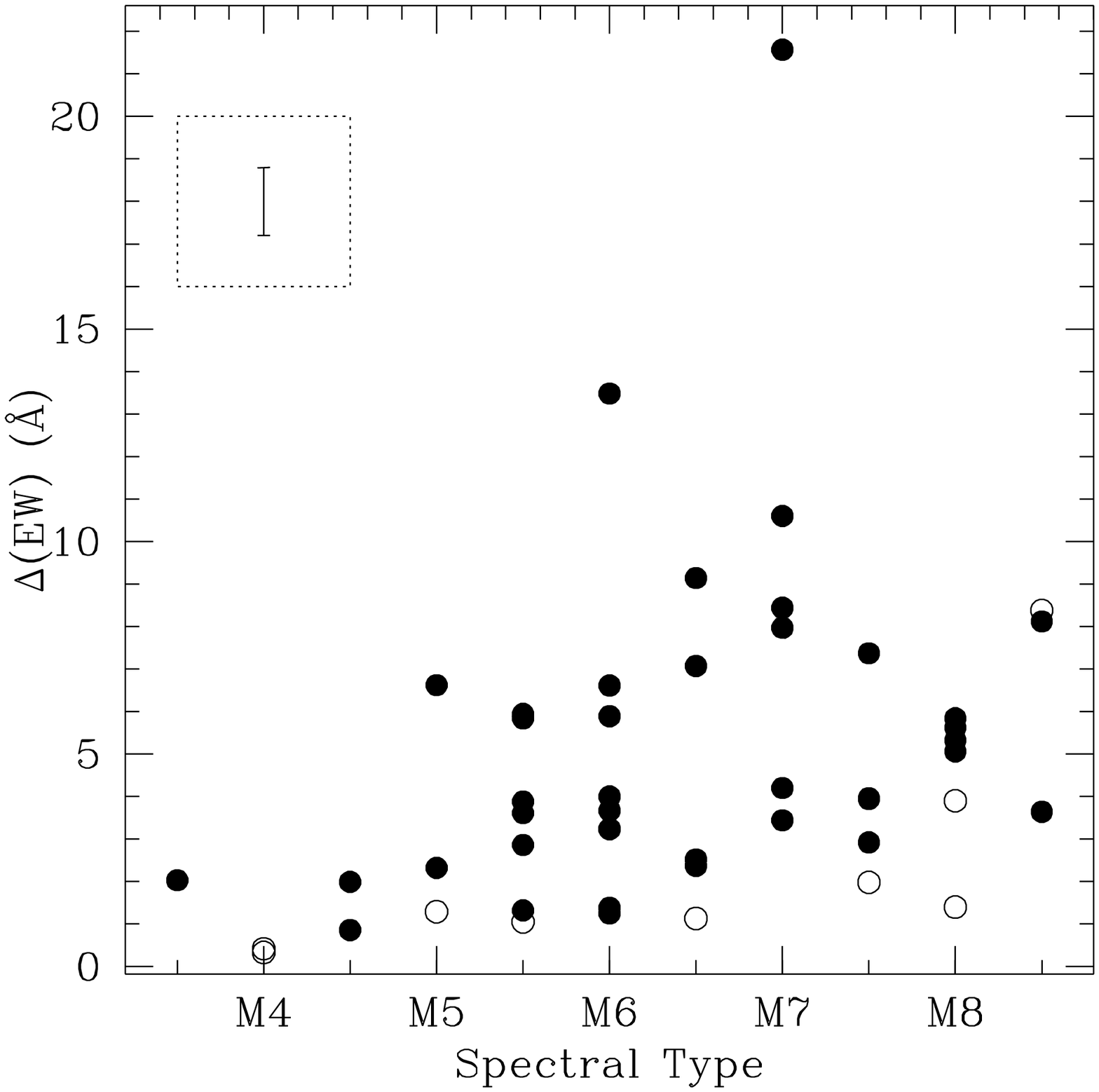,width=2.00in,angle=0}}
\end{minipage}\hfill
\begin{minipage}{2.00in}
\centerline{\psfig{file=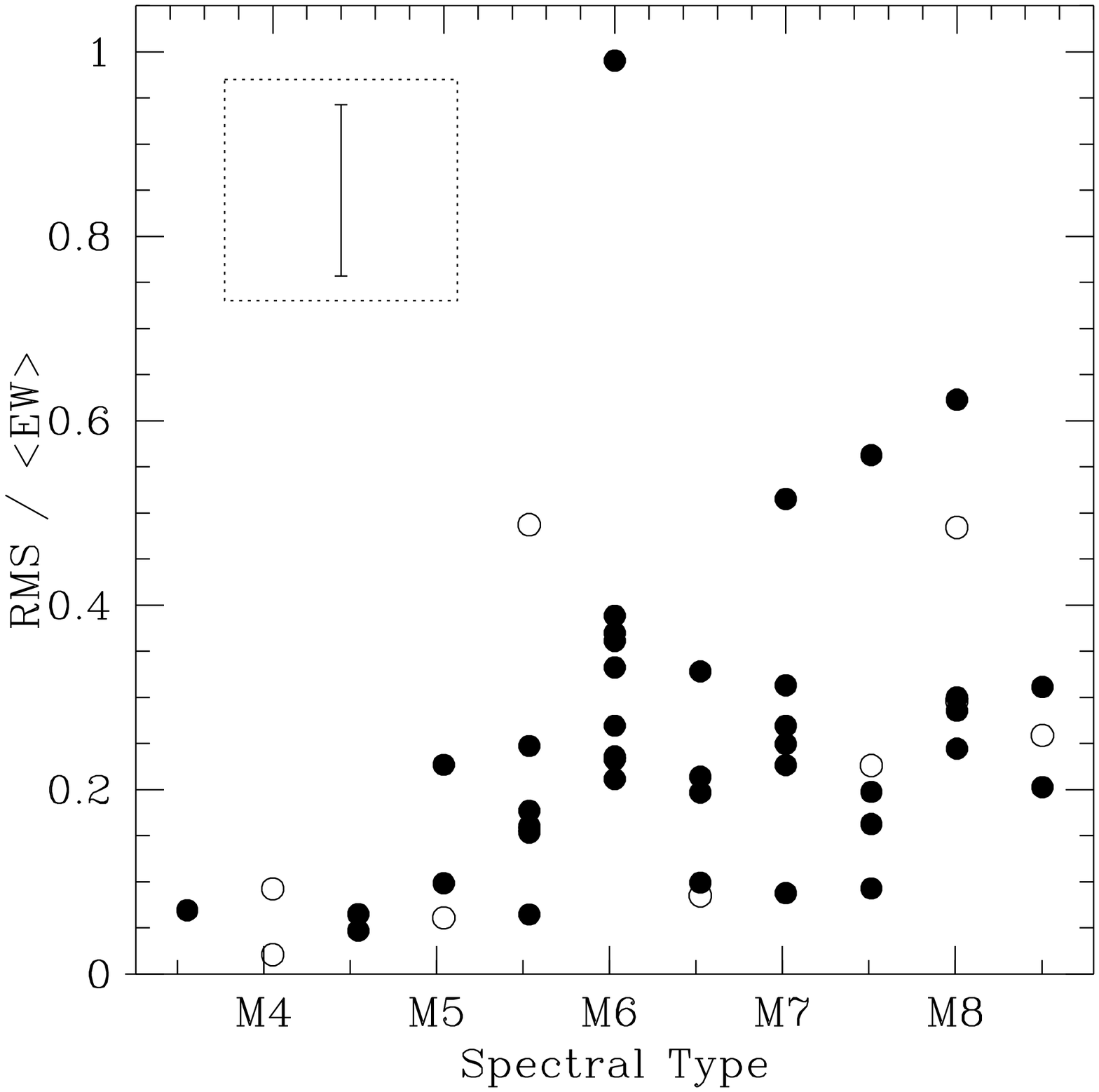,width=2.00in,angle=0}}
\end{minipage}\hfill
\begin{minipage}{2.00in}
\centerline{\psfig{file=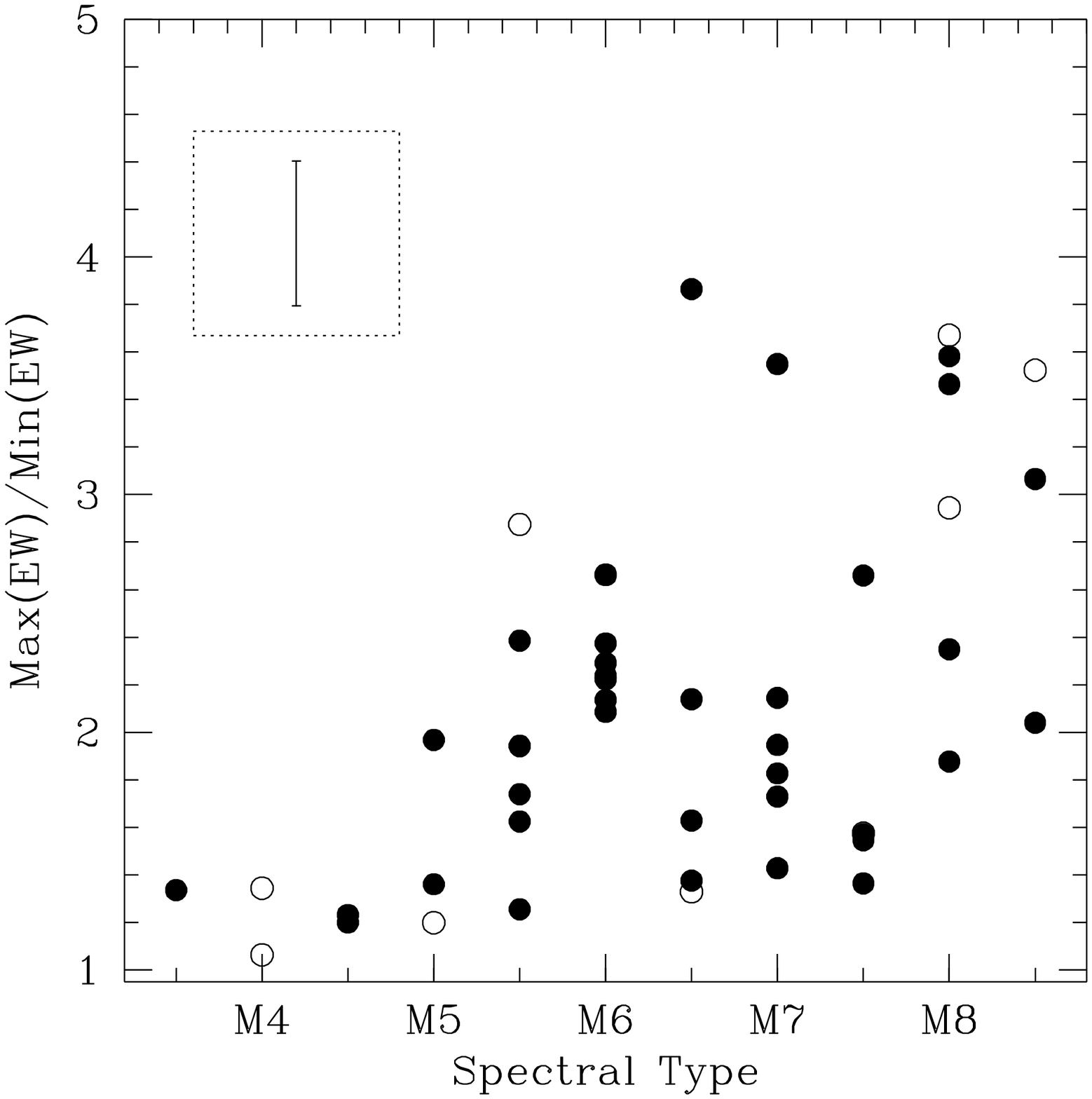,width=2.00in,angle=0}}
\end{minipage}\hfill
\caption{Left: Difference between the maximum and minimum \ha EW,
$\Delta{\rm (EW)}$, plotted against spectral type.  Center: Normalized
RMS of the \ha EW light curves of individual observations, plotted
against spectral type.  Right: Ratio of maximum and minimum \ha EW,
$R{\rm (EW)}$, plotted against spectral type.  In all three plots the
error bar on the upper left show the median errors.  Open circles
designate objects identified as non-varying using our $\chi^2$
criterion (\S\ref{sec:basicres}).
\label{fig:rmsminmaxplt}}
\end{figure*}

We use several indicators to quantify the variability strength.  The
simplest quantity is $\Delta({\rm EW})\equiv \rm Max(EW)-Min(EW)$, the
difference between the maximum and minimum EW values observed for each
object during the course of our observations.  However, since the
conversion between equivalent width and luminosity depends on the
pseudo-continuum brightness (and hence spectral type), this quantity
cannot be easily compared across spectral types.  To account for the
variation in continuum luminosity between spectral types, we use the
ratio of the maximum and minimum EW values, $R({\rm EW})\equiv \rm
Max(EW)/ Min(EW)$, and the RMS normalized by the median equivalent
width, ${\rm RMS(EW)}/\langle {\rm EW}\rangle$.  These quantities are
plotted as a function of spectral type in
Figure~\ref{fig:rmsminmaxplt}.  In all three cases we find a clear
rising trend in variability as a function of spectral type, with an
apparent flattening beyond spectral type M7.

The non-varying objects are included in these and subsequent plots for
completeness.  However, we note that while they may appear to exhibit
significant variability as measured by these metrics, their error bars
are correspondingly larger.  In addition, for several objects we have
more than one observation (Table~\ref{tab:obstable}).  We treat 
the data for these objects as separate observations in order to maintain 
a uniform cadence across our sample, although in Table~\ref{tab:vartable}
we use the combined data.

\begin{figure*}
\plotone{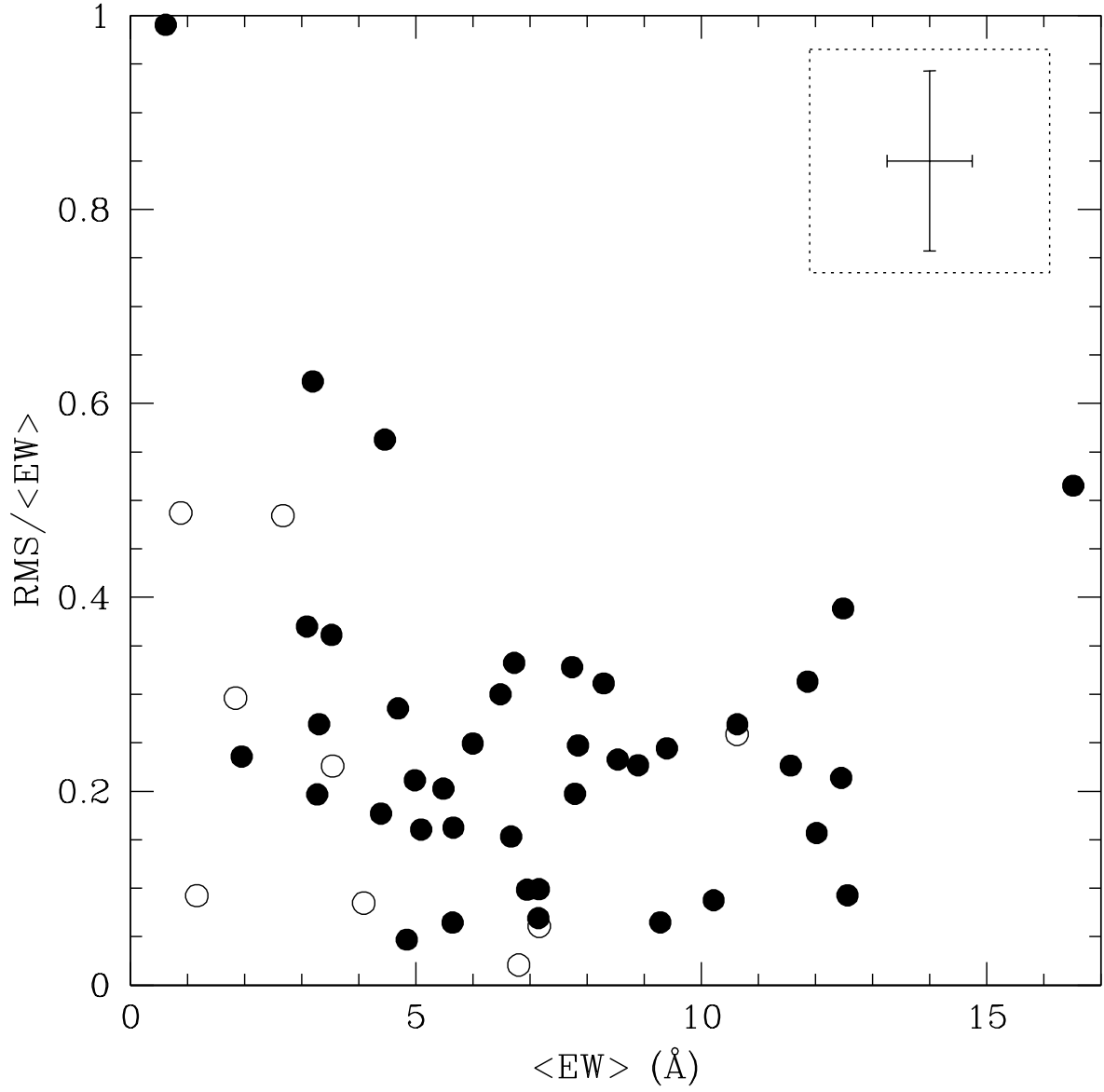}
\caption{Median-normalized RMS of the EW light curves plotted against
the median EW.  The error bars at upper left show the median errors.
Open circles denote objects identified as non-varying through our
$\chi^2$ criterion.
\label{fig:rmsnormed_med}}
\end{figure*}

The distribution of ${\rm RMS(EW)}/\langle {\rm EW}\rangle$ as a
function of $\langle {\rm EW}\rangle$ is shown in
Figure~\ref{fig:rmsnormed_med}.  We find no clear correlation between
the variability and mean equivalent width.  The typical value of ${\rm
RMS(EW)}/\langle {\rm EW}\rangle$ is $\approx 0.25$, and only about
$10\%$ of the objects exceed ${\rm RMS(EW)}/\langle {\rm
EW}\rangle\approx 0.5$.

\begin{figure*}
\plotone{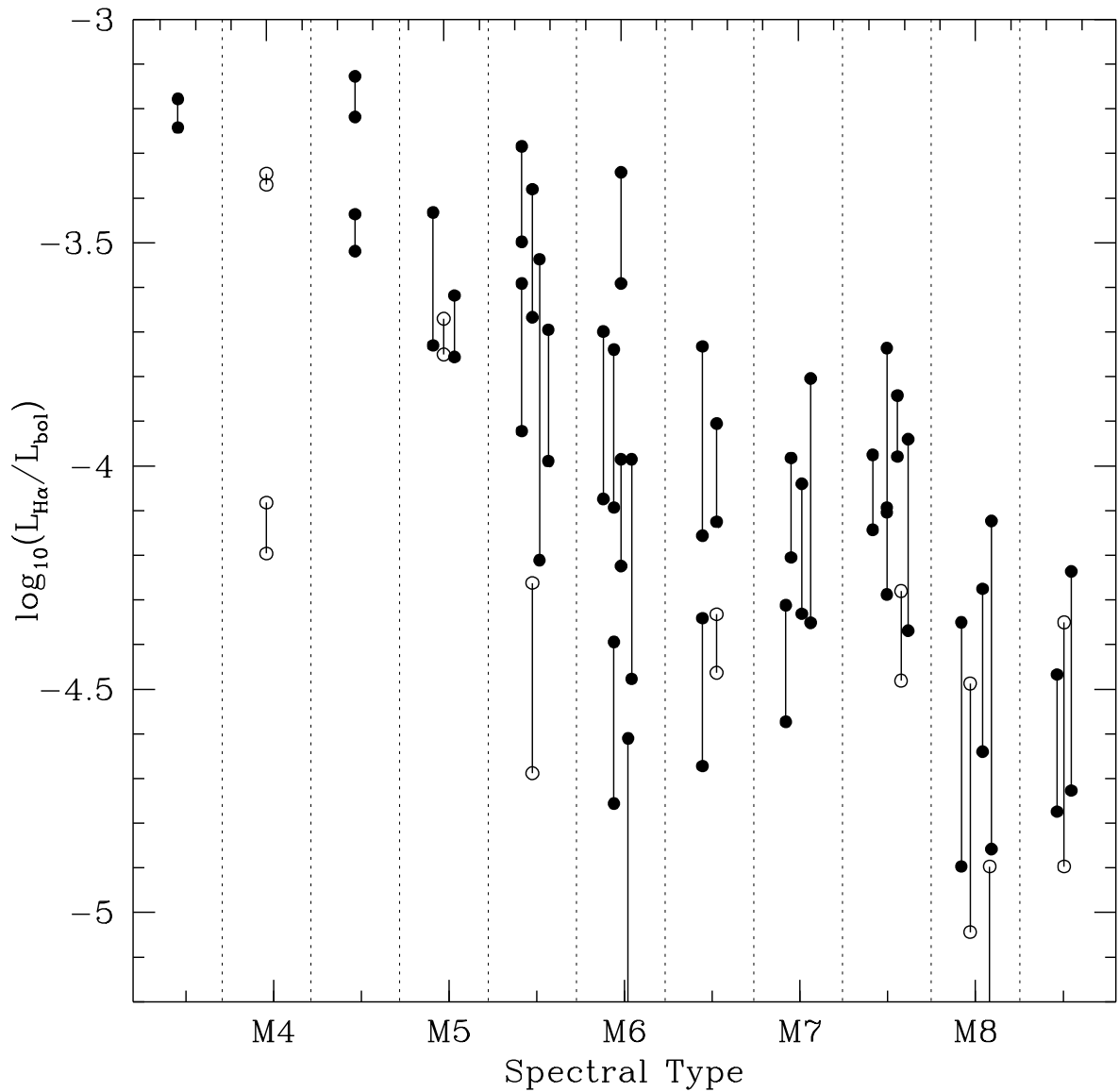}
\caption{Logarithmic ratios of $L_{\rm H\alpha}$ and $L_{\rm bol}$,
plotted against spectral type.  The solid lines connect the maximum
and minimum $L_{\rm H\alpha}/L_{\rm bol}$ values measured for each
object.  Open circles designate objects identified as non-varying
through our $\chi^2$ criterion.  The positions of objects are
displaced horizontally for clarity, and the dashed vertical lines
delineate the spectral types.
\label{fig:lha_spt}}
\end{figure*}

\begin{figure*}
\plotone{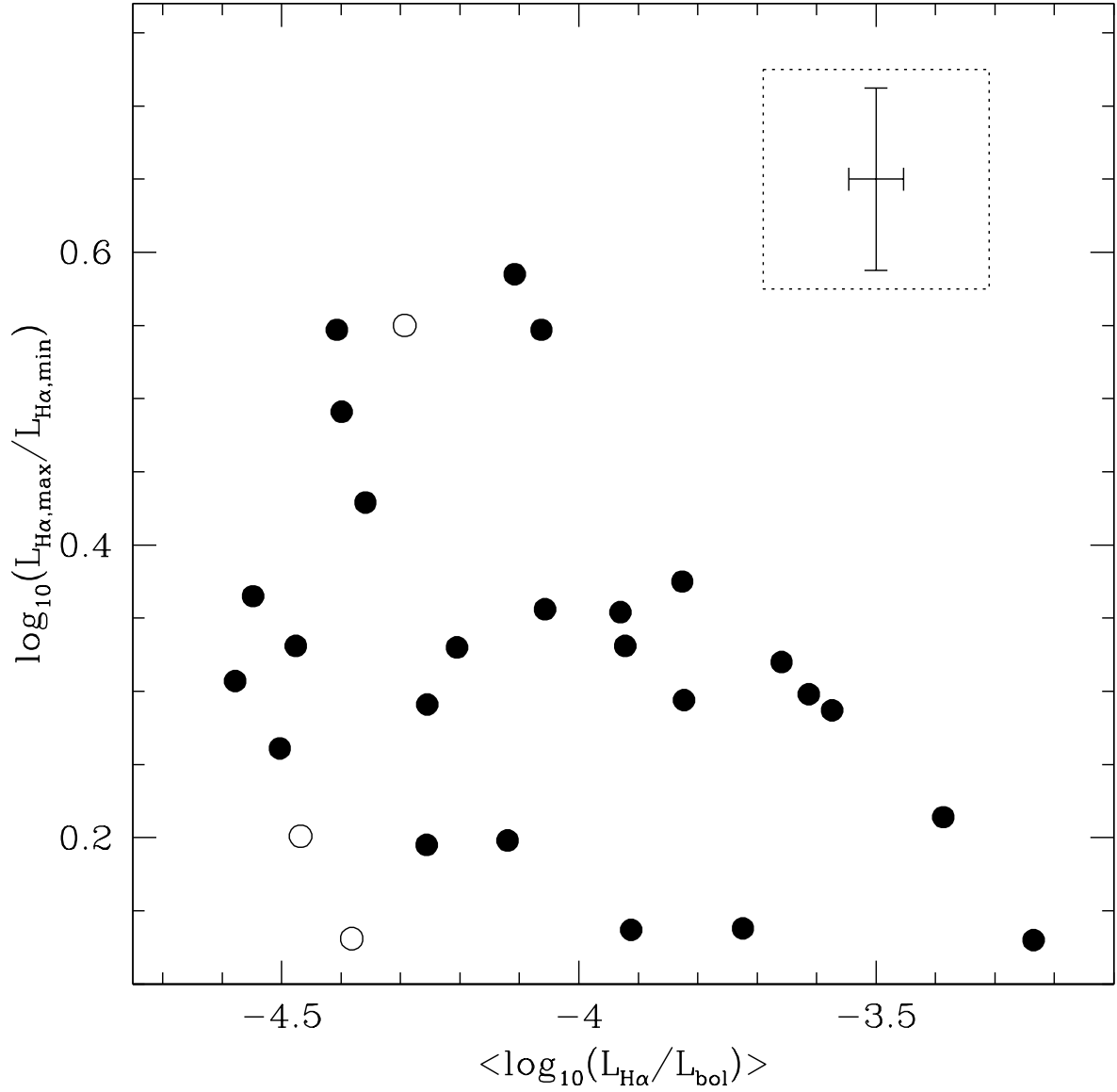}
\caption{Ratio of maximum and minimum \ha luminosity plotted against
median \ha luminosity.  The error bar on the upper right show the
median uncertainty.  Open circles denote objects identified as
non-varying through our $\chi^2$ criterion.
\label{fig:lratio}}
\end{figure*}

Finally, we investigate the variability in terms of \ha luminosity.
The conversion between EW and $\log(L_{\rm H\alpha}/L_{\rm bol})$ is a
function of the spectral type since the continuum luminosity declines
with later spectral types.  Here we adopt the conversion values
(so-called $\chi$ factor values) from \citet{walk04}.  The resulting
mean and maximum values of $\log(L_{\rm H\alpha}/L_{\rm bol})$ are
listed in Table~\ref{tab:vartable}.  In Figure~\ref{fig:lha_spt} we
plot the range of maximum and minimum $\log(L_{\rm H\alpha}/L_{\rm
bol})$ values for each object as a function of spectral type.  We
recover the same overall declining trend in the mean \ha activity as a
function of spectral type demonstrated previously
\citep{hawley00,cr02,lieb03,west04}.  More interestingly, we clearly
find that the level of variability increases with later spectral type,
or equivalently with decreasing $\log(L_{\rm H\alpha}/L_{\rm bol})$.
This result is similarly evident from a comparison of the ratio of
maximum to minimum \ha luminosity to the mean \ha luminosity
(Figure~\ref{fig:lratio}).

\section{Individual Variability Events and Timescales}
\label{sec:var}

To study the distribution of variability amplitudes and timescales in
greater detail, we identify all of the individual variability
``events'' from the light curves; we do not include the non-variable
light curves in this analysis.  Events are defined as EW peaks that
rise by at least $3\sigma$ above the nearest troughs.  For example, in
Figure~\ref{fig:lcurve}e, we identify three events --- a broad event
that lasts from about 10 to 70 min, and two events that last $\lesssim
10$ min at 20 and 35 min.  Similarly, we find four events in
Figure~\ref{fig:lcurve}b --- a decline in the first 15 min, followed
by two short spikes, and finally a gradual rise between 40 and 65 min.
The timescale of each event is defined as the time for the \ha EW to
transition from a trough through a statistically significant peak to
the next trough.  In the case of partial events we use the observed
timescale and amplitudes as lower limits.  We find 71 full events and
27 partial events in the light curves shown in
Figures~\ref{fig:lcurve_all1} and \ref{fig:lcurve_all2}.

\begin{figure*}
\plottwo{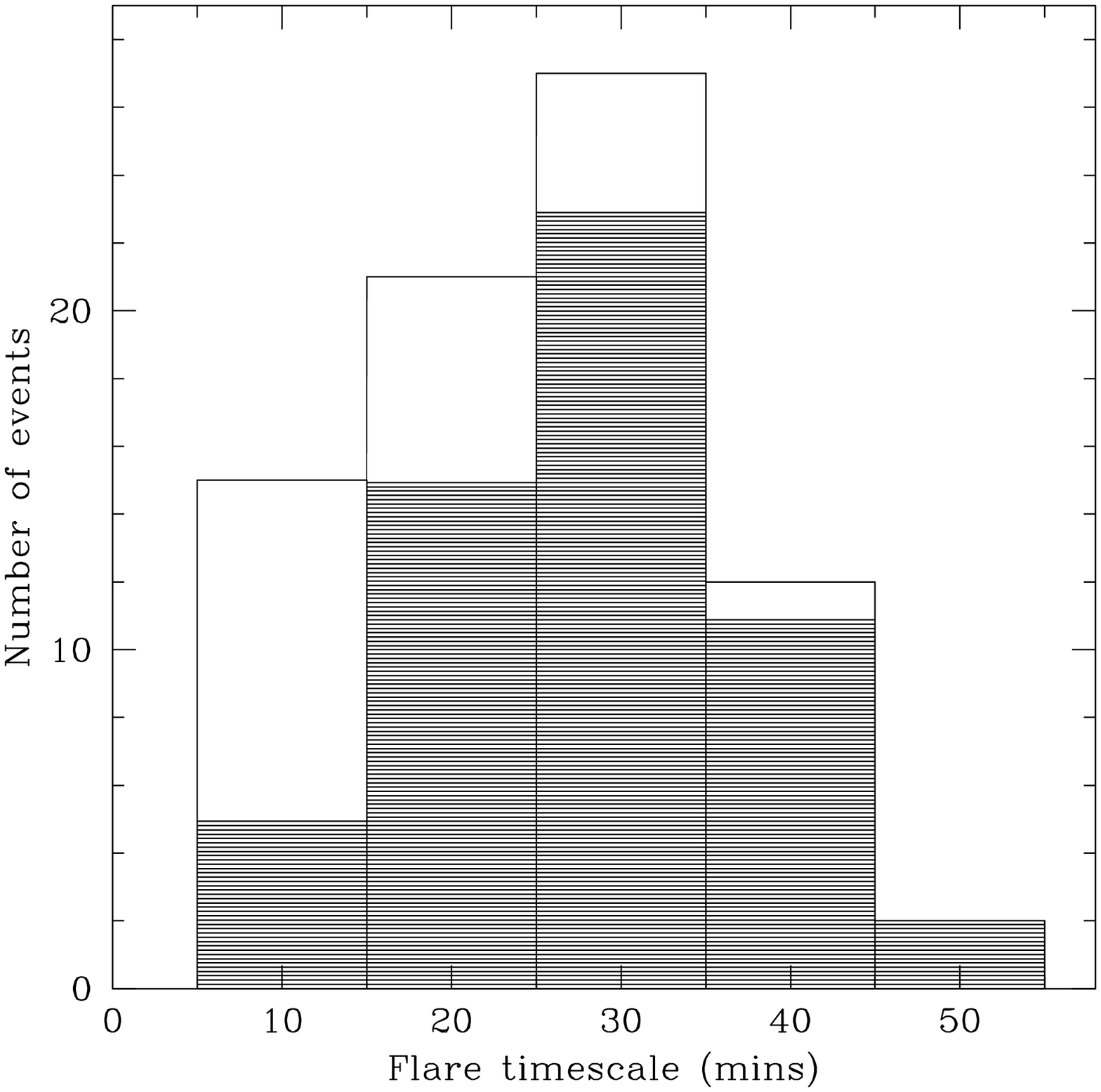}{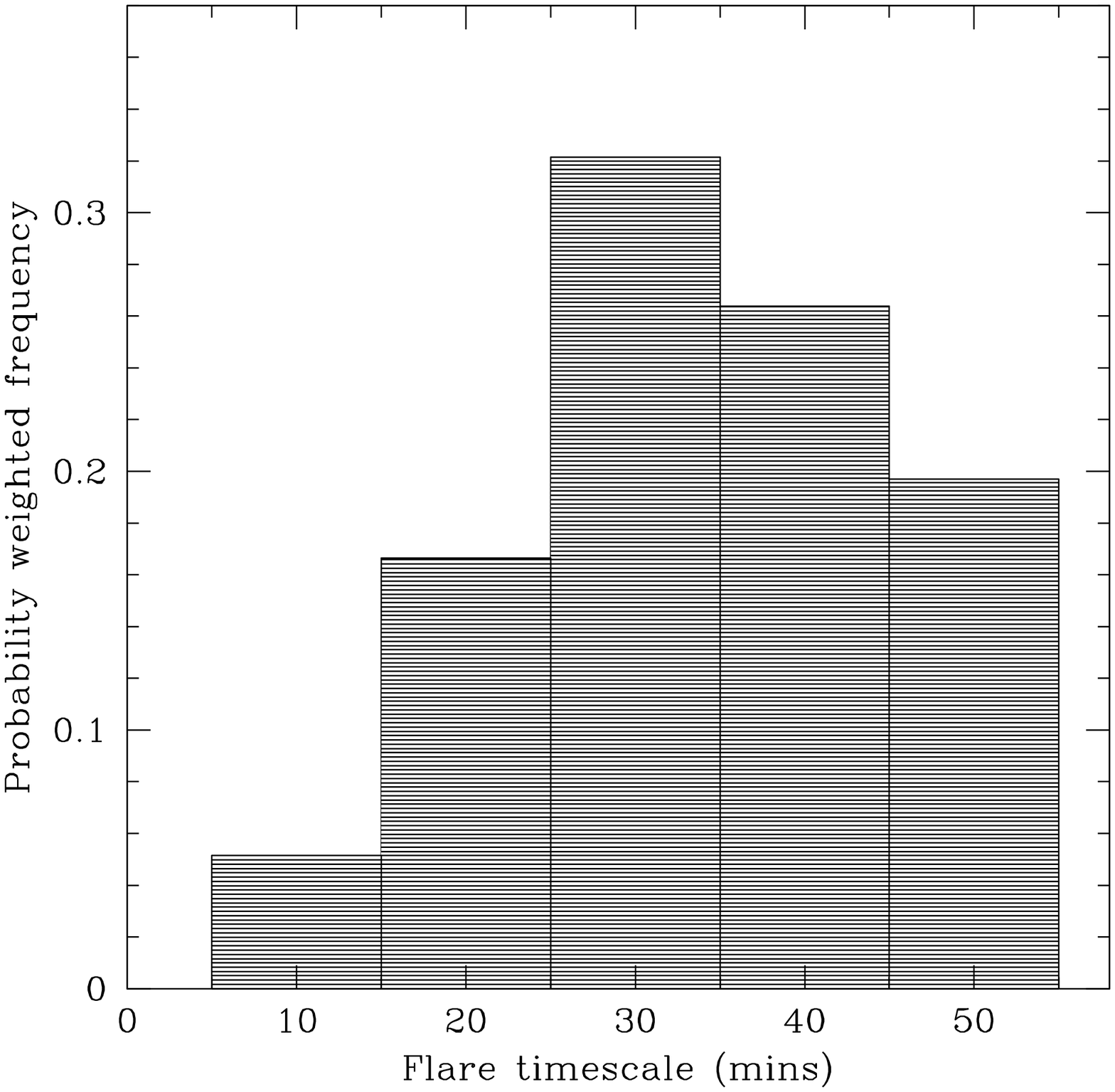}
\caption{Left: Histogram of variability events binned by timescale.
Unshaded regions denote partially observed events.  Right: Frequency
of variability events weighted by the probability of being observed
within a finite observing time of about 1 hr, and with the partial
events uniformly distributed on timescales equal to or longer than
their observed timescale.  The normalization is chosen such that the
total area is unity.
\label{fig:timehist}}
\end{figure*}

A histogram of the variability event timescales is shown in
Figure~\ref{fig:timehist}a.  From the raw event list we find a peak at
about $30$ min, while partial events are naturally clustered at
shorter timescales.  The relatively small number of events with
durations of $\sim 10$ min reflects a real trend since such events can
be easily detected in our light curves.  On the other hand, the
decline beyond 30 min, corresponding to timescales longer than about
one half of an observing sequence, may reflect the diminishing
probability of capturing full events with a duration similar to that
of the observation.  For example, for our typical observations (1 hr
duration with a time resolution of 5 min) the probability of detecting
a full 40 min event is only 5/9 of the probability of detecting a full
20 min event.  To take this effect into account we normalize each full
event bin by its relative probability (Figure~\ref{fig:timehist}b).

In addition, to account for the distribution of partial events we make
the simple assumption that these events have a uniform probability
across all bins with a duration equal to or greater than their
measured duration.  This assumption is somewhat simplistic given the
observed non-uniform trend in the full events, and it is thus likely
to under-estimate the true number of events in the $30-50$ min bins.
However, given the overall number of events this is unlikely to change
our conclusions in a significant way

We show the histogram corrected for event detection probability and
taking into account the partial events in Figure~\ref{fig:timehist}b.
As expected, the trend of increased frequency of events between
timescales of 10 to 30 min becomes significantly more pronounced, with
a factor of 6 times as many events with 30 min duration compared to 10
min duration.  The decline at longer timescales is shallower compared
to the raw data, with a decline of about $30\%$ relative to the 30 min
bin (compared to $60-90\%$ in the raw distribution).  Taking into
account our simple prescription of assigning partial events, it is
likely that the distribution is in fact flat on timescales of $30-60$
min.

\begin{figure*}
\plotone{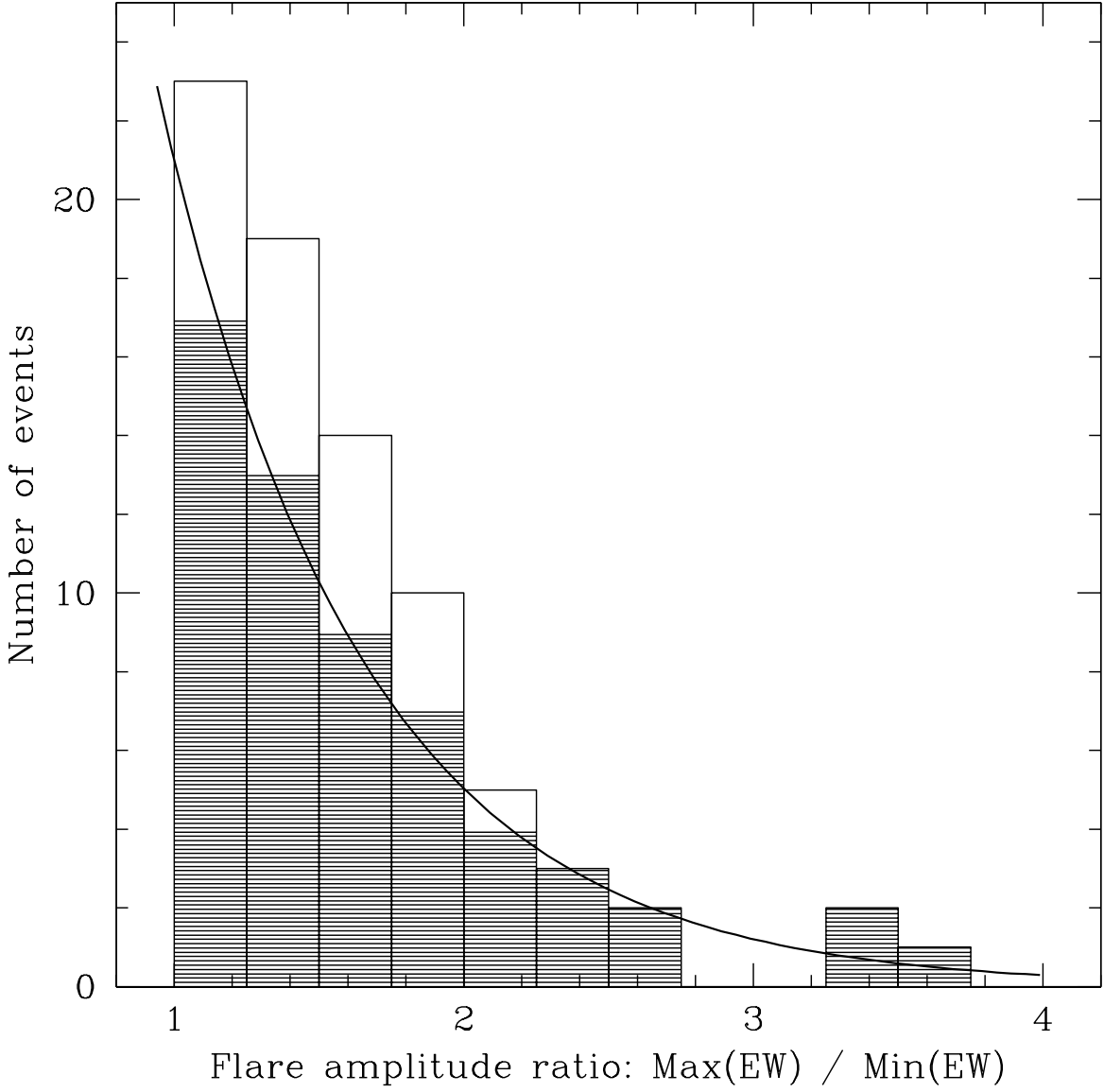}
\caption{Histogram of variability events binned by flare amplitude
ratio.  The solid curve is an exponential distribution with a decay
constant of 0.7.  Unshaded regions denote partially observed events.
\label{fig:amphist}}
\end{figure*}

We next turn to the event amplitude ratios.  We focus on this quantity
since it can be uniformly compared across spectral types.  We plot the
histogram of $R({\rm EW})$ values in Figure~\ref{fig:amphist}.  The
distribution exhibits an exponential decline in the number of events
($N$) as a function of amplitude ratio, with
\begin{equation}
N\propto\exp\left(-\frac{R({\rm EW})-1}{0.7}\right).
\label{eq:ampratio}
\end{equation}
We note that this fit does not include partial events.  To investigate
whether there is a difference in the distribution as a function of
spectral type, we divide the sample into two subsets split at M6 and
repeat the analysis of timescale and amplitude distributions.  We find
the two sub-samples to be indistinguishable, indicating that there no
obvious trend in \ha variability amplitude and timescale for
individual events.

\begin{figure*}
\plotone{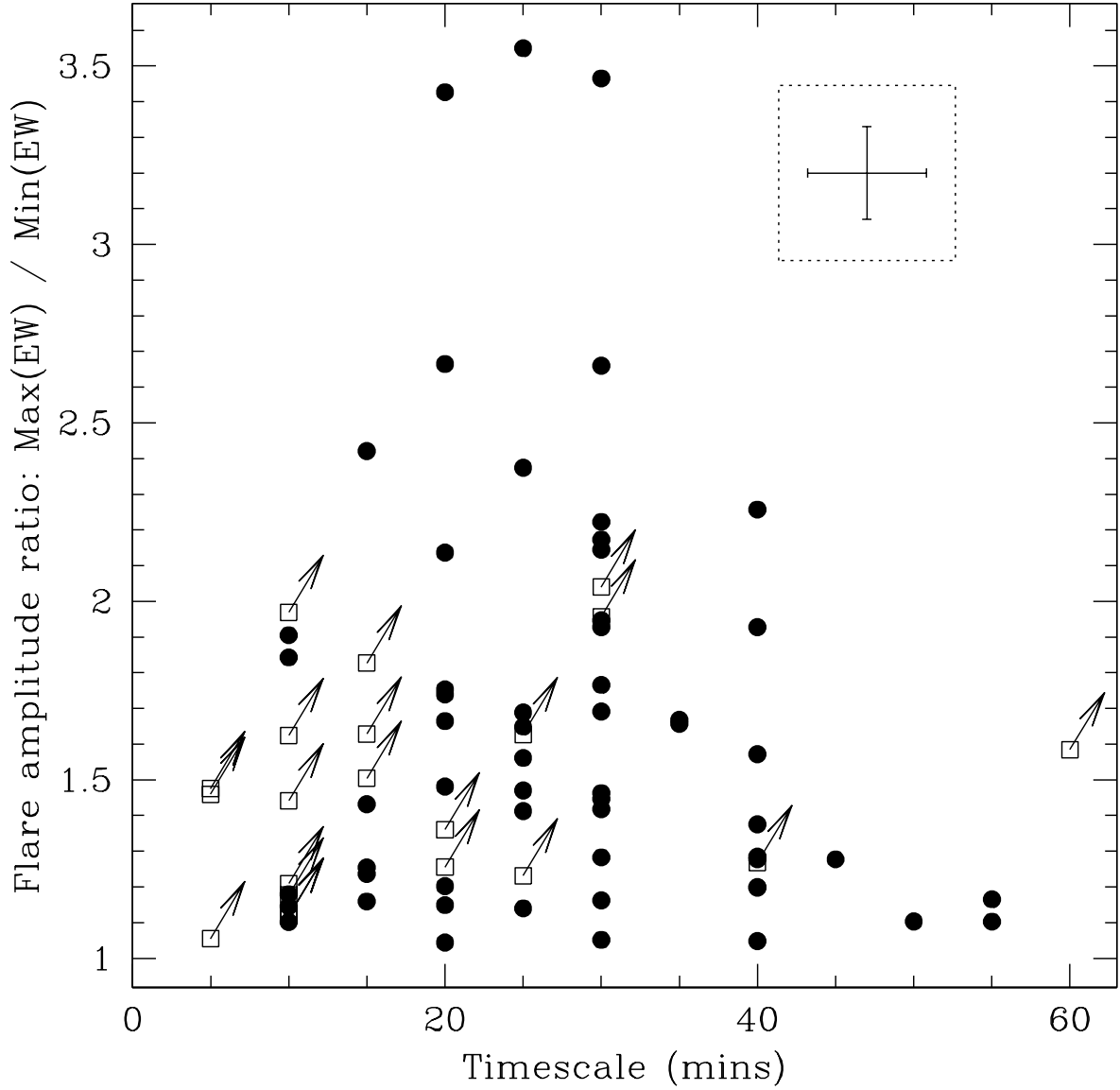}
\caption{Amplitude ratio of variability events plotted against
timescale.  Square boxes with diagonal arrows denote partially
observed events, which are lower limits in both amplitude and
timescale.  The error bars at the upper right show the median errors.
\label{fig:amptime}}
\end{figure*}

Finally, we search for a correlation between the event durations and
amplitudes.  Figure~\ref{fig:amptime} shows the event amplitude ratios
plotted against their durations.  We find that $R ({\rm EW})$
generally increases with increased timescale, but this effect is only
apparent on timescales shorter than about 30 min.  Longer events tend
to have lower amplitudes.  However, partial events are likely to
increase the number of long duration, high amplitude events,
potentially leading to a flatter distribution beyond 30 min.

\section{Discussion and Conclusions}
\label{sec:conc}

We carried out spectroscopic observations of 43 M dwarfs in the range
M3.5--M8.5, for approximately 1 hour each with a time resolutions of
about 5 min.  About 80\% of our targets exhibit statistically
significant H$\alpha$ variability, ranging from a factor of 1.25 to
about 4.  Based on a total on-source exposure time of about 3000 min
we find that the duty cycle of flares with an order of magnitude
increase in brightness is $\lesssim 0.05$ hr$^{-1}$ (95\% confidence
level, assuming Poisson distribution).  This limit is similar to
previous results for individual well-studied objects (e.g., LHS2065;
\citealt{martin01}).

The level of variability for individual objects is found to increase
with later spectral type, with a possible flattening beyond $\sim {\rm
M7}$.  In terms of H$\alpha$ luminosity, we recover the familiar trend
of decreasing $L_{\rm H\alpha}/L_{\rm bol}$ with later spectral type.
More importantly, however, we find that the range of light curve
variability increases with later spectral type, such that M7--M8
objects exhibit a range of about 0.5 dex in $L_{\rm H\alpha}/L_{\rm
bol}$ compared to only 0.15 dex for M4--M5 objects.  This result
indicates that while the traditional definition of ``activity'' (i.e.,
$L_{\rm H\alpha}/L_{\rm bol}$) declines with later spectral type, the
actual fluctuations in activity increase with later spectral type.

The individual light curves exhibit a rich phenomenology, with
activity timescales spanning the full range covered by our
observations, i.e., $\sim 5-60$ min.  We find that fluctuations on
timescales of $\sim 10$ min are significantly less common than those
on $\sim 30$ min timescale, and that longer duration events (at least
to $\sim 1$ hr) are likely to be as common.  The event amplitude
ratios closely follow an exponential distribution with a
characteristic value of $R({\rm EW})-1\approx 0.7$.  Taken at face
value, this means that events with an amplitude ratio of 10 are $\sim
10^5$ times less common than those with a ratio of 2.  Thus, our
observed rate of about 0.1 event per hour with a factor of 2 increase
in EW implies an expected rate of $\sim 10^{-6}$ hr$^{-1}$ for events
with an order of magnitude increase in EW.  The fact that several such
events are published in the literature (e.g., \citealt{martin01})
suggests that major flare events are not drawn from the same
statistical distribution of mild variability events found in our work.
Indeed, \citet{sch07} estimate that events with $\Delta ({\rm EW})
\gtrsim 15$ \AA\ have a duty cycle of $\sim 5\%$ in late-M dwarfs,
orders of magnitude larger than expected from our exponential
distribution (however, $\Delta ({\rm EW})$ depends on spectra type so
a direct comparison is challenging).

Combining the event amplitudes and timescales, we do not find any
clear correlations, although there appears to be a general trend of
declining amplitude for longer duration events.  Such a relation would
be expected if H$\alpha$ events in each spectral type released similar
amounts of total magnetic energy such that $E_{\rm H\alpha}\sim L_{\rm
H\alpha}\times t\propto {\rm EW}\times t\sim {\rm const}$.  On the
other hand, the opposite trend would be expected if H$\alpha$ events
represent the release of magnetic stresses in the chromosphere such
that frequent and/or shorter duration events will have lower
amplitude.  It remains to be seen from observations with longer time
baselines which effect exists, and whether it correlates with spectral
type.

The large and growing samples of nearby M dwarfs are conducive for a
continued systematic investigation of H$\alpha$ emission as a proxy
for magnetic activity.  Clearly, H$\alpha$ variability at the level of
$\lesssim 2$ is prevalent in the bulk of H$\alpha$-emitting mid- and
late-M dwarfs on timescales of $\sim 0.1-1$ hr.  Future observations
will need to address three primary questions:
\begin{enumerate}
\item Are the durations and amplitudes of H$\alpha$ events correlated?
If so, directly or inversely?
\item Are large flares drawn from the same distribution as small
events?
\item Are the prevalence and properties of events/flares correlated
with age or rotation velocity across the M spectral type range?
\end{enumerate}
To address these questions we will continue to pursue observations of
larger samples, as well as longer time baseline observations of
individual active objects.

\acknowledgements This research has made use of the SIMBAD database,
operated at CDS, Strasbourg, France, and of NASA's Astrophysics Data
System Bibliographic Services.

\input{manualbib.tex}
\input{table1.tex}

\input{table2.tex}

\clearpage

\clearpage


\clearpage

\clearpage

\clearpage

\clearpage

\clearpage

\clearpage

\end{document}

%% file: table1.tex
\clearpage
\begin{deluxetable}{llccccccl}
\tablewidth{0pt}
\tabletypesize{\scriptsize}
\tablecolumns{9}
\tabcolsep0.04in\footnotesize
\tablecaption{Source Properties and Log of Observations \label{tab:obstable}}
\tablehead{
\colhead{Source}     & 
\colhead{Other}      & 
\colhead{Sp.~Type}   & 
\colhead{Date\,\tablenotemark{a}} &
\colhead{Exposures}  &
\colhead{log($L_{\rm bol}/L_{\sun}$)\,\tablenotemark{b}}&
\colhead{\it{d}}     &
\colhead{$v\sin i$}  &
\colhead{Ref.\,\tablenotemark{c}} \\ 
\colhead{}           & 
\colhead{}           & 
\colhead{}           & 
\colhead{(UT)}       &   
\colhead{(s)}        & 
\colhead{}           &
\colhead{(pc)}       &
\colhead{(km s$^{-1}$)} & 
\colhead{}           
}
\startdata
G 99-049                  &   GJ 3379     &  M3.5 &  Mar 15 23:58:23   & 12$\times$300 & -1.63  &      5.4  &     7.4      &        1, \it{8}  \\ 
LHS 1723                  &   \nodata     &  M4   &  Mar 14 00:01:34   & 11$\times$300 & -1.92  &      6.1  &    $<$3.2      &       2, \it{8} \\
L 449-1                   &   \nodata     &  M4   &  Mar 14 01:00:14   & 11$\times$300 &\nodata   &      5.7  &  \nodata    &        3         \\
GJ 1224                   &   \nodata     &  M4.5 &  Mar 14 08:29:12   & 14$\times$300 &    -2.36  &      7.5  &   $ <$5.6   &      2, \it{8} \\ 
GL 285                    &   V* YZ CMi   &  M4.5 &  Mar 15 01:10:51   & 13$\times$300 &    -1.52  &      6.2  &      6.5    &      2, \it{8} \\
2MASSW J1013426-275958    &   \nodata     &  M5   &  Mar 14 04:28:58   & 6$\times$600  &  \nodata  &  \nodata  &  \nodata   &     4, \it{8}   \\
GJ 1156                   &   V* GL Vir   &  M5   &  Mar 14 05:57:50   & 12$\times$300 &    -2.30  &      6.5  &      9.2    &    2, \it{8}  \\
GJ 1154A                  &   \nodata     &  M5   &  Mar 15 03:41:37   & 18$\times$300 &  \nodata  &      8.5  &      5.2   &    2, \it{8}    \\
DENIS-P J213422.2-431610  &   \nodata     &  M5.5 &  Sep 12 04:12:02   & 4$\times$300  &  \nodata  &     14.6  &  \nodata   &      5          \\
2MASS J02591181+0046468   &   \nodata     &  M5.5 &  Sep 14 04:48:39   & 11$\times$300 &    -3.61  &       29  &  \nodata   &      6          \\
2MASS J02534448-7959133   &   \nodata     &  M5.5 &  Sep 14 06:21:38   & 13$\times$300 &    -3.44  &     17.2  &  \nodata   &      7          \\
2MASS J00244419-2708242   &   GJ 2005     &  M5.5 &  Sep 14 07:42:30   & 12$\times$300 &    -2.59  &      7.5  &    9.0    &      8, \it{8} \\
                          &               &       &  Sep 15 03:20:25   & 10$\times$300 &         &             &           &                \\
                          &               &       &  Sep 15 06:14:05   & 38$\times$300 &         &             &         &                 \\
2MASS J00045753-1709369   &   \nodata     &  M5.5 &  Sep 15 00:52:58   & 9$\times$300  &    -3.31  &     14.9 & \nodata &        5         \\
2MASS J20021341-5425558   &   \nodata     &  M5.5 &  Sep 12 02:51:06   & 11$\times$300 &    -3.55  &     17.2 & \nodata   &      7         \\
LP 844-25                 &   LHS 2067    &  M6   &  Mar 14 03:19:25   & 6$\times$600  &    -3.90  &     25.1 & \nodata    &      9       \\
2MASS J16142520-0251009   &   LP 624-54   &  M6   &  Mar 17 08:34:42   & 15$\times$300 &    -3.42  &     14.6  & \nodata   &        7       \\
                          &               &       &  Sep 13 23:47:29   & 10$\times$300 &           &         &            &                \\
2MASS J21322975-0511585   &   NLTT 51488  &  M6   &  Sep 14 23:52:34   & 6$\times$300  &    -3.47  &   18.5  & \nodata   &          7      \\
2MASS J23373831-1250277   &   NLTT 57439  &  M6   &  Sep 14 01:47:33   & 16$\times$300 &    -3.49  &  19.2  & \nodata     &         10     \\
2MASSW J1012065-304926    &   \nodata     &  M6   &  Mar 17 01:13:30   & 6$\times$600  &  \nodata  &  \nodata  & \nodata  &          4      \\
LP 731-47                 &   \nodata     &  M6   &  Mar 17 04:29:30   & 12$\times$300 &  \nodata  &     20.5  &  11.0    &    7, \it{8}   \\
2MASS J23155449-0627462   &   NLTT 56283  &  M6   &  Sep 16 00:40:42   & 9$\times$300  &  -3.37  &   17.7  & \nodata      &          11     \\
2MASS J20424514-0500193   &   NLTT 49734  &  M6.5 &  Sep 13 03:32:18   & 12$\times$300 &    -3.51  &   15.5  & \nodata      &         12    \\
GJ 3622                   &  \nodata      &  M6.5 &  Mar 15 05:25:04   & 12$\times$300 &  \nodata  &      4.5  &   3.0     &  1, \it{8}     \\
2MASS J05023867-3227500   &   \nodata     &  M6.5 &  Mar 16 23:56:31   & 12$\times$300 &  -3.89  &   25.1  & \nodata      &         10      \\
2MASS J02141251-0357434   &   LHS 1363    &  M6.5 &  Sep 13 09:01:56   & 10$\times$300 &    -3.11  &   10.1  & \nodata      &       10      \\
                          &               &       &  Sep 14 03:25:22   & 4$\times$300  &         &             &           &               \\
2MASS J10031918-0105079   &   LHS 5165    &  M7   &  Mar 14 02:08:39   & 6$\times$600  &    -3.84  &  23.1  & \nodata    &          10     \\
2MASS J13092185-2330350   &   \nodata     &  M7   &  Mar 14 07:11:58   & 6$\times$600  &    -3.62  &     13.3  &   7.0  &        4, \it{13} \\
2MASSW J1032136-420856    &   \nodata     &  M7   &  Mar 15 02:42:22   & 6$\times$600  & \nodata  & \nodata  & \nodata    &        4       \\
2MASSW J1420544-361322    &   \nodata     &  M7   &  Mar 15 06:37:15   & 8$\times$300  &  \nodata  &  \nodata  & \nodata  &         4       \\
2MASS J09522188-1924319   &   \nodata     &  M7.5 &  Mar 16 02:26:22   & 10$\times$450 &    -3.68  &  \nodata  &   6.0     &     13, \it{13} \\
                          &               &       &  Mar 17 23:45:15   & 8$\times$600  &          &           &            &                \\
2MASS J04291842-3123568   &  \nodata      &  M7.5 &  Sep 12 07:12:29   & 12$\times$300 &    -3.26  &  11.4  &  \nodata    &          14    \\
2MASS J23062928-0502285   &  \nodata      &  M7.5 &  Sep 13 05:11:27   & 6$\times$600  &    -3.46  &  11.0  &  \nodata   &           10    \\
2MASS J03313025-3042383   &   NLTT 11163  &  M7.5 &  Sep 13 06:28:07   & 6$\times$600  &    -3.46  &  12.1  &  \nodata   &           10    \\
2MASS J04351612-1606574   &   NLTT 13580  &  M7.5 &  Sep 13 07:44:12   & 12$\times$300 &    -3.09  &   8.6  &  \nodata    &          15    \\
2MASS J06572547-4019134   &   \nodata     &  M7.5 &  Mar 16 01:22:06   & 6$\times$600  &    -4.01  &  22.7  &  \nodata  &            10    \\
2MASS J05173766-3349027   &   \nodata     &  M8   &  Sep 12 08:27:48   & 9$\times$600  &    -3.71  &  14.7  &  \nodata   &           10    \\
2MASS J19165762+0509021   &   VB 10       &  M8   &  Sep 13 00:54:33   & 12$\times$300 &    -2.88  &   5.7  &  6.5       &      8, \it{8} \\
2MASS J22062280-2047058   &   \nodata     &  M8   &  Sep 14 00:35:30   & 6$\times$600  &    -3.88  &     18.2  &  22.0  &      10, \it{8}  \\
                          &               &       &  Sep 15 23:39:08   & 6$\times$600  &           &          &          &                \\
2MASS J02484100-1651216   &  \nodata      &  M8   &  Sep 14 09:10:55   & 4$\times$600  &    -3.94  &  16.2  &  \nodata   &         10     \\
2MASS J20370715-1137569   &  \nodata      &  M8   &  Sep 16 23:39:45   & 5$\times$600  &    -3.85  &  16.8  &  \nodata   &        10      \\
2MASS J22264440-7503425   &   \nodata     &  M8.5 &  Sep 12 01:26:16   & 6$\times$600  &    -3.87  &  16.5  &  \nodata    &       5       \\
                          &               &       &  Sep 15 04:35:01   & 6$\times$600  &          &         &             &               \\
2MASS J03061159-3647528   &   \nodata     &  M8.5 &  Sep 12 04:50:23   & 5$\times$600  &    -3.61  &  11.3  &  \nodata    &       5       \\
2MASS J23312174-2749500   &   \nodata     &  M8.5 &  Sep 13 02:13:20   & 6$\times$600  &    -3.61  &  11.6  &  \nodata    &       5       \\
\enddata
\tablenotetext{a}{All observations were carried out in the year 2007.}
\tablenotetext{b}{Bolometric luminosities were derived using
bolometric corrections on the $J$ and $K$ magnitudes listed in the
{\sc Simbad} database, using the fits described in \citet{wilking99}.}
\tablenotetext{c}{References for spectral types are in standard font;
references for rotational velocities are in italics: [1]
\citet{henry94}; [2] \citet{d98}; [3] \citet{scholz05}; [4]
\citet{gizis02}; [5] \citet{crifo05}; [6] \citet{boch05}; [7]
\citet{pb06}; [8] \citet{mb03}; [9] \citet{rg05}; [10] \citet{cruz03};
[11] \citet{2scholz05}; [12] \citet{reid04}; [13] \citet{reid02}; [14]
\citet{sch07}; [15] \citet{lod05}.}
\end{deluxetable}

%% file: table2.tex
\clearpage
\begin{deluxetable}{lcccccccccc}
\rotate
\tablewidth{0pt}
\tabletypesize{\scriptsize}
\tablecolumns{11}
\tabcolsep0.04in\footnotesize
\tablecaption{H$\alpha$ Variability \label{tab:vartable}}
\tablehead{
\colhead{Source}       & 
\colhead{Sp.~Type}     &  
\colhead{Published EW} & 
\colhead{Ref.\tablenotemark{a}} &
\colhead{$\langle {\rm EW}\rangle$\tablenotemark{b}} & 
\colhead{Min(EW)}      & 
\colhead{Max(EW)}      &
\colhead{$\Delta {\rm (EW)}$\tablenotemark{c}} & 
\colhead{RMS(EW)\tablenotemark{d}}  &
\colhead{Mean}         & 
\colhead{Max}          \\
\colhead{}             &
\colhead{}             &
\colhead{}             &
\colhead{}             &
\colhead{}             &
\colhead{}             &
\colhead{}             &
\colhead{}             &
\colhead{}             &
\colhead{log($L_{\rm H\alpha}/L_{\rm bol}$)} & 
\colhead{log($L_{\rm H\alpha}/L_{\rm bol}$)} \\
\colhead{}             &
\colhead{}             &
\colhead{(\AA)}        &
\colhead{}             &
\colhead{(\AA)}        &
\colhead{(\AA)}        &
\colhead{(\AA)}        &
\colhead{(\AA)}        &
\colhead{(\AA)}        &
\colhead{}             &
\colhead{}             
}
\startdata
G 99-049                 &  M3.5  & 2.9     & 1       & 7.1 $\pm$ 0.1 & 7.0 $\pm$ 0.1 & 8.1 $\pm$ 0.1 & 1.1 $\pm$ 0.1 & 0.5 $\pm$ 0.1   &  -3.229  &  -3.178 \\
LHS 1723                 &  M4    & 0.9     & 1       & 1.2 $\pm$ 0.2 & 1.0 $\pm$ 0.1 & 1.3 $\pm$ 0.1 & 0.3 $\pm$ 0.2 & 0.1 $\pm$ 0.1(C)&  -4.155  &  -4.082 \\ 
L 449-1                  &  M4    & \nodata & \nodata & 6.8 $\pm$ 0.2 & 6.7 $\pm$ 0.1 & 7.1 $\pm$ 0.1 & 0.4 $\pm$ 0.2 & 0.1 $\pm$ 0.1(C)&  -3.357  &  -3.345 \\
GJ 1224                  &  M4.5  & 2.3     & 1       & 4.8 $\pm$ 0.2 & 4.3 $\pm$ 0.2 & 5.2 $\pm$ 0.2 & 0.9 $\pm$ 0.2 & 0.2 $\pm$ 0.2   &  -3.480  &  -3.436 \\  
GL 285                   &  M4.5  & 9.5     & 1       & 9.3 $\pm$ 0.1 & 8.6 $\pm$ 0.1 & 10.6$\pm$ 0.1 & 2.0 $\pm$ 0.1 & 0.6 $\pm$ 0.1   &  -3.179  &  -3.152 \\ 
2MASSW J1013426-275958   &  M5    & 6.6     & 2       & 7.2 $\pm$ 0.4 & 6.5 $\pm$ 0.4 & 7.8 $\pm$ 0.3 & 1.3 $\pm$ 0.5 & 0.4 $\pm$ 0.3(C)&  -3.711  &  -3.670 \\  
GJ 1156                  &  M5    & 4.4     & 1       & 6.9 $\pm$ 0.3 & 6.4 $\pm$ 0.2 & 8.8 $\pm$ 0.2 & 2.3 $\pm$ 0.3 & 0.7 $\pm$ 0.2   &  -3.705  &  -3.618 \\  
GJ 1154A                 &  M5    & 4.3     & 1       & 8.9 $\pm$ 0.3 & 6.8 $\pm$ 0.2 & 13.5$\pm$ 0.2 & 6.6 $\pm$ 0.2 & 2.0 $\pm$ 0.2   &  -3.589  &  -3.448 \\
DENIS-P J213422.2-431610 &  M5.5  & \nodata & \nodata & 0.9 $\pm$ 0.5 & 0.6 $\pm$ 0.3 & 1.6 $\pm$ 0.3 & 1.1 $\pm$ 0.4 & 0.4 $\pm$ 0.4(C)&  -4.466  &  -4.262 \\
2MASS J02591181+0046468  &  M5.5  &  14.5   & 3       &12.0 $\pm$ 0.8 & 9.3 $\pm$ 0.7 & 15.2$\pm$ 0.4 & 5.8 $\pm$ 0.8 & 1.9 $\pm$ 0.6   &  -3.387  &  -3.290 \\  
2MASS J02534448-7959133  &  M5.5  & 12.4    & 4       & 7.8 $\pm$ 0.3 & 6.3 $\pm$ 0.4 & 12.2$\pm$ 0.6 & 5.9 $\pm$ 0.7 & 1.9 $\pm$ 0.5   &  -3.502  &  -3.387 \\
2MASS J00244419-2708242  &  M5.5  & 3.6     & 1       & 4.0 $\pm$ 4.5 & 1.8 $\pm$ 4.5 & 8.5 $\pm$ 0.3 & 6.7 $\pm$ 4.5 & 1.1 $\pm$ 0.7   &  -3.734  &  -3.537  \\
2MASS J00045753-1709369  &  M5.5  & \nodata & \nodata & 4.4 $\pm$ 0.4 & 3.0 $\pm$ 0.3 & 5.9 $\pm$ 0.4 & 2.9 $\pm$ 0.5 & 0.8 $\pm$ 0.4   &  -3.813  &  -3.695 \\
2MASS J20021341-5425558  &  M5.5  & 7.6     & 4       & 1.9 $\pm$ 0.4 & 3.5 $\pm$ 0.8 & 7.5 $\pm$ 0.5 & 4.0 $\pm$ 0.9 & 1.3 $\pm$ 0.6   &  -3.785  &  -3.591 \\
LP 844-25                &  M6    & \nodata &\nodata  & 6.7 $\pm$ 1.2 & 1.0 $\pm$ 0.3 & 2.3 $\pm$ 0.3 & 1.2 $\pm$ 0.4 & 0.5 $\pm$ 0.3   &  -4.610  &  -4.394 \\
2MASS J16142520-0251009  &  M6    & 4.2     & 4       & 8.5 $\pm$ 0.9 & 1.9 $\pm$ 0.7 & 5.9 $\pm$ 0.6 & 3.9 $\pm$ 0.9 & 1.0 $\pm$ 0.6   &  -4.187  &  -3.985  \\
2MASS J21322975-0511585  &  M6    & 1.1     & 4       & 3.5 $\pm$ 0.8 & 0.1 $\pm$ 0.5 & 1.4 $\pm$ 0.5 & 1.4 $\pm$ 0.7 & 0.6 $\pm$ 0.5   &  -5.057  &  -4.610 \\
2MASS J23373831-1250277  &  M6    & \nodata & \nodata & 3.1 $\pm$ 0.7 & 14.6$\pm$ 0.5 & 25.9$\pm$ 0.6 & 11.3$\pm$ 0.8 & 4.9 $\pm$ 0.5   &  -3.500  &  -3.358 \\
2MASSW J1012065-304926   &  M6    & 5.6     & 2       &12.5 $\pm$ 0.7 & 4.6 $\pm$ 0.7 & 10.4$\pm$ 0.4 & 5.8 $\pm$ 0.8 & 2.2 $\pm$ 0.7   &  -3.904  &  -3.756 \\
LP 731-47                &  M6    & 6.4     & 4       & 0.6 $\pm$ 0.7 & 4.8 $\pm$ 0.7 & 11.4$\pm$ 1.0 & 6.6 $\pm$ 1.2 & 2.0 $\pm$ 0.7   &  -3.847  &  -3.714 \\
2MASS J23155449-0627462  &  M6    & 1.1     & 5       & 5.0 $\pm$ 0.4 & 3.4 $\pm$ 0.9 & 5.9 $\pm$ 0.7 & 2.5 $\pm$ 1.2 & 1.1 $\pm$ 0.8   &  -4.122  &  -3.985 \\
2MASS J20424514-0500193  &  M6.5  & \nodata & \nodata & 4.1 $\pm$ 0.4 & 2.1 $\pm$ 0.5 & 4.5 $\pm$ 0.4 & 2.4 $\pm$ 0.6 & 0.6 $\pm$ 0.5   &  -4.451  &  -4.341 \\
GJ 3622                  &  M6.5  & 1.9     & 1       & 7.7 $\pm$ 0.8 & 3.4 $\pm$ 0.3 & 4.6 $\pm$ 0.3 & 1.1 $\pm$ 0.4 & 0.3 $\pm$ 0.3(C)&  -4.393  &  -4.332 \\
2MASS J05023867-3227500  &  M6.5  & \nodata & \nodata & 3.3 $\pm$ 0.7 & 7.4 $\pm$ 0.5 & 12.3$\pm$ 2.5 & 4.9 $\pm$ 2.5 & 2.5 $\pm$ 1.0   &  -4.050  &  -3.915 \\
2MASS J02141251-0357434  &  M6.5  & \nodata & \nodata & 7.2 $\pm$ 0.4 & 6.9 $\pm$ 0.3 & 18.3$\pm$ 0.5 & 11.5$\pm$ 0.6 & 3.7 $\pm$ 0.4   &  -4.012  &  -3.732 \\
2MASS J10031918-0105079  &  M7    & \nodata & \nodata &10.6 $\pm$ 1.2 & 8.9 $\pm$ 0.9 & 17.4$\pm$ 0.7 & 8.4 $\pm$ 1.1 & 2.9 $\pm$ 0.8   &  -4.201  &  -4.050 \\
2MASS J13092185-2330350  &  M7    & 6.7     & 2       & 6.0 $\pm$ 0.9 & 5.1 $\pm$ 0.6 & 9.3 $\pm$ 0.5 & 4.2 $\pm$ 0.8 & 1.5 $\pm$ 0.6   &  -4.383  &  -4.312 \\
2MASSW J1032136-420856   &  M7    & 11.8    & 2       &11.9 $\pm$ 1.3 & 11.9$\pm$ 1.3 & 19.9$\pm$ 1.4 & 8.0 $\pm$ 1.9 & 3.7 $\pm$ 1.4   &  -4.201  &  -4.002 \\
2MASSW J1420544-361322   &  M7    & 7.0     & 2       &16.5 $\pm$ 1.2 & 8.5 $\pm$ 0.6 & 30.0$\pm$ 0.5 & 21.6$\pm$ 0.8 & 8.5 $\pm$ 0.8   &  -4.050  &  -3.804 \\
2MASS J09522188-1924319  &  M7.5  & 11.3    & 6       & 4.5 $\pm$ 2.1 & 8.1 $\pm$ 0.6 & 18.9$\pm$ 0.3 & 10.8$\pm$ 0.7 & 3.1 $\pm$ 0.4   &  -3.940  &  -3.736  \\
2MASS J04291842-3123568  &  M7.5  & 15.9    & 7       &10.9 $\pm$ 0.7 & 10.8$\pm$ 0.4 & 14.8$\pm$ 0.5 & 3.9 $\pm$ 0.6 & 1.2 $\pm$ 0.5   &  -3.933  &  -3.866 \\
2MASS J23062928-0502285  &  M7.5  & 2.8     & 7       &12.6 $\pm$ 0.5 & 3.4 $\pm$ 0.7 & 5.4 $\pm$ 0.8 & 2.0 $\pm$ 1.1 & 0.8 $\pm$ 0.7(C)&  -4.379  &  -4.280 \\
2MASS J03313025-3042383  &  M7.5  & 7.6     & 7       & 3.5 $\pm$ 0.9 & 7.4 $\pm$ 0.6 & 10.9$\pm$ 0.6 & 3.5 $\pm$ 0.8 & 1.5 $\pm$ 0.7   &  -4.068  &  -4.012 \\
2MASS J04351612-1606574  &  M7.5  & 3.7     & 6       & 7.8 $\pm$ 1.0 & 5.3 $\pm$ 0.8 & 8.3 $\pm$ 0.4 & 2.9 $\pm$ 0.8 & 0.9 $\pm$ 0.6   &  -4.206  &  -4.093 \\
2MASS J06572547-4019134  &  M7.5  & \nodata & \nodata & 5.7 $\pm$ 0.8 & 4.4 $\pm$ 1.2 & 11.8$\pm$ 1.7 & 7.4 $\pm$ 2.1 & 2.5 $\pm$ 1.8   &  -4.180  &  -3.971 \\
2MASS J05173766-3349027  &  M8    & 8.1     & 4       & 2.7 $\pm$ 1.6 & 1.5 $\pm$ 1.2 & 5.4 $\pm$ 1.1 & 3.9 $\pm$ 1.6 & 1.3 $\pm$ 1.1(C)&  -4.688  &  -4.487 \\
2MASS J19165762+0509021  &  M8    & 5.6     & 1       & 4.7 $\pm$ 0.8 & 3.8 $\pm$ 0.5 & 8.8 $\pm$ 0.5 & 5.1 $\pm$ 0.7 & 1.3 $\pm$ 0.5   &  -4.479  &  -4.275 \\
2MASS J22062280-2047058  &  M8    & 5.1     & 8       & 3.6 $\pm$ 1.3 & 2.3 $\pm$ 0.9 & 12.5$\pm$ 1.2 & 10.2$\pm$ 1.5 & 2.9 $\pm$ 1.1   &  -4.356  &  -4.123  \\
2MASS J02484100-1651216  &  M8    & 7.9     & 7       & 6.5 $\pm$ 1.2 & 2.1 $\pm$ 1.3 & 7.4 $\pm$ 1.0 & 5.3 $\pm$ 1.7 & 1.9 $\pm$ 1.2   &  -4.538  &  -4.350 \\
2MASS J20370715-1137569  &  M8    & 6.2     & 7       & 1.8 $\pm$ 1.3 & 0.7 $\pm$ 0.7 & 2.1 $\pm$ 1.4 & 1.4 $\pm$ 1.5 & 0.5 $\pm$ 1.2(C)&  -5.016  &  -4.897 \\
2MASS J22264440-7503425  &  M8.5  & 3.2     & 7       & 0.6 $\pm$ 2.4 & 3.3 $\pm$ 1.1 & 11.7$\pm$ 1.5 & 8.4 $\pm$ 1.9 & 2.7 $\pm$ 1.7(C)&  -4.538  &  -4.350  \\
2MASS J03061159-3647528  &  M8.5  & 0.5     & 6       & 8.3 $\pm$ 0.7 & 3.9 $\pm$ 0.8 & 12.1$\pm$ 0.7 & 8.1 $\pm$ 1.1 & 2.6 $\pm$ 0.8   &  -4.421  &  -4.239 \\
2MASS J23312174-2749500  &  M8.5  & 5.1     & 6       & 5.5 $\pm$ 1.0 & 3.5 $\pm$ 0.9 & 7.1 $\pm$ 0.7 & 3.6 $\pm$ 1.1 & 1.1 $\pm$ 0.9   &  -4.594  &  -4.467 \\
\enddata
\tablenotetext{a}{References for the published EWs: 
(1) {\citet{mb03}}
(2) {\citet{gizis02}}
(3) {\citet{boch05}}
(4) {\citet{pb06}}
(5) {\citet{reid03}}
(6) {\citet{lod05}}
(7) {\citet{sch07}}
(8) {\citet{reid02}} }
\tablenotetext{b}{Median EW.}
\tablenotetext{c}{$\Delta \mathrm{(EW)} \equiv \mathrm{Max(EW)}-\mathrm{Min(EW)}$}
\tablenotetext{d}{`(C)' denotes that the object's EW was constant within errors during
the observations (see discussion in Section~\ref{sec:basicres}). 
Note that while 2M2226-7503
had two separate observations, 
both lightcurves were found to be consistent with a constant EW.}
\end{deluxetable}